\documentclass[twocolumn,showpacs,preprintnumbers,amsmath,amssymb]{revtex4}

\usepackage{graphicx}
\usepackage{dcolumn}
\usepackage{bm}

\begin{document}

\title{Non-destructive, dynamic detectors for Bose-Einstein 
condensates}

\author{J. E. Lye}
\author{J. J. Hope}
\author{J. D. Close}
 \email{john.close@anu.edu.au}

\affiliation{
Australian Centre for Quantum Atom Optics, \\
Australian National University, ACT 0200, Australia}

\date{\today}

\begin{abstract}
We propose and analyze a series of non-destructive, dynamic detectors 
for
Bose-Einstein condensates based on photo-detectors operating at the
shot noise limit.  These detectors are compatible with real time 
feedback to 
the condensate.  The signal to noise ratio of different detection
schemes are compared subject to the constraint of minimal heating due
to photon absorption and spontaneous emission.  This constraint leads
to different optimal operating points for interference-based schemes. 
We find the somewhat counter-intuitive result that without the
presence of a cavity, interferometry causes as much destruction as
absorption for optically thin clouds.  For optically thick clouds,
cavity-free interferometry is superior to absorption, but it still
cannot be made arbitrarily non-destructive .  We propose a
cavity-based measurement of atomic density which can in principle be
made arbitrarily non-destructive for a given signal to noise ratio.
\end{abstract}


\maketitle

\section{\label{sec:intro}Introduction}

This paper presents a comprehensive analysis of non-destructive,
dynamic measurement schemes of Bose Einstein condensates (BEC) in both
interferometric and non-interferometric configurations.  
The dynamic nature of these detectors is essential if they are to be 
used for
feedback to the condensate. 
Optical detection of the condensate causes heating
through photon absorption and spontaneous emission, and this prompted
the development of nondestructive techniques that detect the
phase shift imparted on a laser beam rather than the absorption of 
that beam
\cite{Ketterle1996,Ketterle1999,Lye1999,Aspect1999,Hulet1997,Milburn1998,Walker2001}.

It is therefore necessary to compare the signal to noise
ratio (SNR) achievable by each technique {\it for a given
absorption rate}.  This constraint changes both the
optimum operating conditions for many techniques as well as the 
optimal
choice of detection scheme in different parameter regimes.  We find 
that no 
current technique can be made arbitrarily sensitive for fixed 
absorption, 
and propose a new detection scheme based on an optical cavity that 
has a 
sensitivity under this criterion that scales with the square root of 
the finesse.

These detection schemes are based on fast photo-detectors operating 
at the
shot noise limit.  In contrast to all existing techniques based on CCD
cameras, these schemes allow for real-time density measurements with
high temporal resolution that are appropriate for the implementation
of feedback to the condensate.  This feedback will initially allow
mode-locking of the BEC, and eventually allow control of its quantum
state that will in turn determine the properties of an atom laser beam
\cite{Wiseman2002, Thomsen2001}.  The development of these detectors 
and
feedback is important if we are to realize the pumped atom laser and
through it, the full potential of quantum atom optics.

We find, contrary to popular belief, that the SNR in interferometric
measurements cannot be increased arbitrarily by increasing laser power
and increasing detuning from atomic resonance.  Further, we find that 
for 
thin clouds subject to a fixed heating by the probe beam,
fluorescence measurements can be more sensitive than single-pass 
interferometric measurements such as those that have been performed.  
This is
surprising given that fluorescence is based on the destructive
phenomenon of absorption and spontaneous emission whereas
interferometry is sensitive to the phase shift of the forward
scattered photons, suggesting interferometry would always be the less
invasive technique.  For each technique, we derive expressions for 
the 
minimum observable change in the column density of the condensate as 
a function of
heating and bandwidth for optically thick and thin
clouds.  The techniques we discuss are compatible
with optical amplitude/phase and spatial squeezing allowing sub-shot
noise and sub-diffraction limited resolution in future
implementations \cite{Treps2002,Mckenzie2002}.  

\begin{figure}
\includegraphics{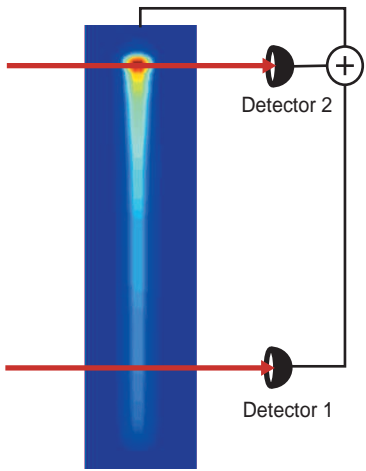}
\caption{\label{fig:figthought} An atom laser with detection and
feedback.  Complete stabilization of an atom laser may require
detection of the condensate directly (detector 2) as well as 
detection of
the atom
beam itself (detector 1).}
\end{figure}

The requirements on dynamic detectors are best illustrated by
consideration of the Gedanken experiment sketched in 
Fig.~\ref{fig:figthought}.
Here, we consider an atom laser beam produced
by coherent outcoupling from a condensate that may be pumped,
although it will be unpumped in early
investigations \cite{laser1,laser2,laser3,laser4,laser5}.  The atom 
beam
and condensate are
probed by light beams, and information regarding the noise and
fluctuations of the condensate and the atom beam is fed back to the
condensate.  The requirements on the design of the two detectors shown
are quite different.  For the dynamic detection of the atom laser beam
with detector 1, there is no non-destructive criterion
\cite{Esslinger2002}.  This can be
seen by analogy with the detection of photons from a laser beam.
Nothing could be more destructive to an optical beam than a
photodiode, photomultiplier or CCD camera.  The photons are destroyed
and an electron is excited to a new state and recorded.  For example,
atoms outcoupled from a metastable helium BEC can
be counted using a multi-channel plate with good time resolution and
high quantum efficiency\cite{Leduc2001,Aspect2001}.  No such detector
exists for neutral ground
state atoms.  The design of such a detector will be the subject of a
future paper.

It might be thought that feedback from detector 1 would be sufficient
to stabilize the atom laser.  Any classical noise brought
about by motion of the condensate in the trap could probably be
compensated by feedback from the atom laser beam flux.  In the absence
of classical noise, Wiseman and Thomsen have recently concluded that a
pumped atom laser will have a linewidth dominated by the effect of the
atomic interaction energy, which turns fluctuations in the condensate
number into fluctuations in the condensate phase \cite{Thomsen2001}.  
They
further conclude that feedback from the atom laser beam flux will not
improve the linewidth, and suggest using dispersive imaging of the
condensate and feeding back to the phase of the condensate via the
trap bias or via a far-detuned laser beam.  This role is fulfilled by
detector 2 and the feedback loop shown in Fig. 1.  Detector 2 is the
more difficult of the two detectors to design and implement, and we
concentrate on the non-destructive dynamic detection of the condensate
in this paper.

The optimization of a measurement scheme, whether it be
interferometry, absorption or fluorescence is strongly influenced by 
the
restrictions imposed by the physics of the system being probed.
Gravity-wave interferometers, for example, are limited by laser power
and saturation of the detectors.  Optimization of a shot-noise limited
(SNL)
phase measurement with these restrictions requires operation near a
dark port with equal power in the interferometer arms.  This design is
used in all current gravity wave detectors under
development \cite{Gray1995,Caves1981,Loudon1981,McClelland1993}.  In
contrast, the non-destructive criterion for the probing of a
condensate that we apply in this paper leads to unbalanced powers in
the interferometer arms.  As another example, optimization of the SNR
for the non-destructive interferometric detection of
condensates while holding the ratio of probe to local oscillator power
constant leads to increasing signal to noise with increasing detuning
\cite{Lye1999, Aspect1999}.
In many of the designs we discuss here, the local oscillator is passed
around the condensate.  Although total absorption by the BEC is fixed,
total power in the interferometer is not, and the SNR
becomes independent of detuning from resonance at least for optically
thin clouds.  The following sections contain a detailed analysis of
shot-noise limited measurements optimized for the non-destructive, dynamic 
detection of Bose Einstein condensates.

\section{\label{sec:Bigab}Dynamic, non-destructive absorption and 
fluorescence.}

\subsection{\label{sec:ab}Absorption}

In a single beam absorption measurement, a probe with incident power
$P_{p0}$ passes through the atoms.  The probe beam receives a phase
shift $\phi$ and is partially absorbed, with the power transmitted
described by $P_{pt}=P_{p0}e^{-k}$.  Both the phase shift and
absorption coefficient, $k$, scale linearly with the column density,
$\tilde{n}_{0} =\int n~dz$, the density integrated along the beam
direction.  Stabilization of the BEC via feedback requires detecting a
small fluctuating component of the column density, which may be on a
large, slowly changing column density background.  We explicitly
define both components to correctly optimize the detection in the
limits of both optically thick and thin clouds.

\begin{eqnarray}
    \label{eq:2bits}
\tilde{n}_{0} = \tilde{n}+\delta \tilde{n} \sin(\omega_{p}t)
\nonumber\\
\phi = \phi_{p}+\delta \phi_{p} \sin(\omega_{p}t)
\nonumber\\
k = k_{p}+\delta k_{p} \sin(\omega_{p}t)
\end{eqnarray}

Ultimately, we want a sensitive detector that can detect very small
fluctuations.  We assume $~\delta \phi_{p},~\delta k_{p}\ll 1$.  
Setting
$\delta \tilde{n} = \tilde{n}$ describes the specific case of
detecting the full BEC, in the limit of thin clouds.

The absorption and phase shift are given by:
\begin{equation}
    \label{eq:ab}
     k_{p}=\tilde{n} \sigma_{0}\frac{1}{1+\Delta^{2}} \textnormal{ 
and }
     \delta k_{p} =\delta \tilde{n} \sigma_{0}\frac{1}{1+\Delta^{2}}
\end{equation} \begin{equation}
    \label{eq:phase}
    \phi_{p}=\frac{\tilde{n} \sigma_{0}}{2}\frac{\Delta}{1+\Delta^{2}}
    \textnormal{ and }
\delta \phi_{p}=\frac{\delta \tilde{n}
\sigma_{0}}{2}\frac{\Delta}{1+\Delta^{2}}
\end{equation}
where $\sigma_{0}=\frac{3\lambda^{2}}{2\pi}$ is the resonant 
absorption
cross-section,
 assuming we are probing a closed transition.  The detuning given in 
units
of half atomic linewidth,
is $\Delta\equiv\frac{\omega-\omega_{0}}{\gamma /2}$.

The optical power transmitted through the BEC is detected on a 
phase-insensitive photodetector. 
The detector responsivity, $\rho$, relates the incident optical power 
to 
the current produced by the photodiode. 
The quantum efficiency, $\eta$, is related to the responsivity through
$\rho=\frac{\eta e}{h\nu}$.

\begin{equation}
    \label{eq:isigab}
i=\rho P_{p0}(e^{-k_{p}}-e^{-k_{p}}\delta k_{p} \sin(\omega_{p} t))
\end{equation}

The desired signal is the AC component of $i$, which is selected
with an appropriate filter on the current.  The resulting 
RMS 
signal, $\sqrt{\langle i_{ac}^{2}\rangle}$, is:
\begin{equation}
    \label{eq:sigab}
i_{sig}=\frac{\rho}{\sqrt{2}}P_{p0}e^{-k_{p}}\delta k_{p}
\end{equation}

We assume the noise is dominated by shot noise from the laser, which
translates to current noise on the photodetector \cite{Yariv1985},
\begin{equation}
    \label{eq:shot}
    i_{shot}=\sqrt{2eB\langle i \rangle},
\end{equation}
where $B$ is the bandwidth of the detection system.

\begin{equation}
    \label{eq:shotab}
i_{shot}=\sqrt{2eB \rho P_{p0}e^{-k_{p}}}
\end{equation}

Taking the ratio of Eqs.  (\ref{eq:sigab}) and (\ref{eq:shotab}), and
setting $\delta k_{p}=\frac{\delta \tilde{n}}{\tilde{n}} k_{p}$, gives
the SNR for a SNL dynamic absorption measurement:

\begin{equation}
\frac{S}{N}=\sqrt{\frac{\eta}{4Bh\nu}P_{p0}e^{-k_{p}}}k_{p}\frac{\delta
\tilde{n}}{\tilde{n}}
\end{equation}

\subsubsection{Adding a non-destructive criterion}

A continuous `non-destructive' measurement sets an upper limit on the
average power absorbed, $\langle P_{ab} \rangle=\langle P_{p0}-P_{pt}
\rangle$, by the BEC.
\begin{eqnarray}
    \label{eq:nd}
 \langle P_{ab} \rangle&=& P_{pt}\langle e^{k_{p}+\delta
     k_{p}\sin(\omega_{p}t)}-1\rangle \nonumber\\
 &=& P_{pt}(e^{k_{p}}-1) ~~~~~~~~~~~~~~~~~~\textnormal{or} \nonumber 
\\ 
 &=& P_{p0}(1-e^{-k_{p}})
\end{eqnarray}
For an optically thin cloud at steady state, this absorbed power is
equal to the power emitted by the process of spontaneous emission. 
This can be converted directly to the number of photons emitted per
atom every second.  It is the recoil from these spontaneous emission
events which will destroy and dephase the BEC. As the BEC becomes
optically thick, reabsorption of the emitted photons becomes an issue,
and the total number of recoils caused by the absorption of a single
photon from the probe laser beam will depend on the mean free path of
the photons in the medium as well as the geometry of the trapped
atoms.  In this regime, every photon absorbed by the BEC will do far
more damage than a single recoil event, and the restriction on the
amount of power absorbed by the BEC for a non-destructive measurement
will need to be more stringent.

Including the non-destructive criterion above in the SNR:

\begin{equation}
\frac{S}{N}=\sqrt{\frac{\eta\langle
P_{ab}\rangle}{4Bh\nu}}\frac{\delta
\tilde{n}}{\tilde{n}}\sqrt{\frac{e^{-k_{p}}k_{p}^{2}} {1-e^{-k_{p}}}}
\end{equation}

\begin{figure}
\includegraphics{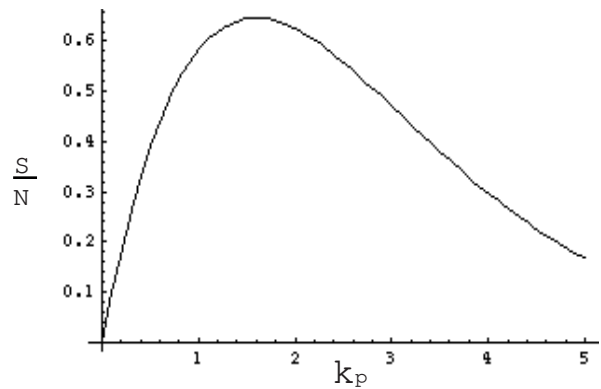}
\caption{\label{fig:figk} The normalized signal to noise ratio for a
non-destructive SNL absorption measurement as a function of the
absorption coefficient.  There is a clear optimum at $k_{p}=1.6$,
corresponding to approximately 80\% of the probe power absorbed.}
\end{figure}

We optimize the function $\frac{e^{-k_{p}}k_{p}^{2}}{1-e^{-k_{p}}}$,
shown in Figure \ref{fig:figk}.  The
function reaches a maximum value of 0.65 when $k_{p}=1.6$, which
corresponds to approximately 80\% of the power absorbed.  At larger 
absorption, the total amount of power absorbed
increases, but the sensitivity of the absorption to small fluctuations
decreases as the BEC becomes optically thick.

The maximum absorption possible occurs on resonance, when
$k_{p}=\tilde{n}\sigma_{0}$.  If $\tilde{n}\sigma_{0} > 1.6$, the
optimum value $k_{p}=1.6$ can be chosen by detuning the probe beam
appropriately.  If $\tilde{n}\sigma_{0} < 1.6$, $k_{p}$ should be set
to its maximum value, $\tilde{n}\sigma_{0}$, by putting the probe on
resonance.  This leads to two different optimum SNR equations in the
limits of optically thin and thick clouds:

\begin{equation}
    \label{eq:snabthick}
\frac{S}{N}_{thick}=\sqrt{\frac{\eta\langle
P_{ab}\rangle}{2.5Bh\nu}}\frac{\delta \tilde{n}}{\tilde{n}}
\end{equation}
\begin{equation}
    \label{eq:snabthin}
\frac{S}{N}_{thin}=\sqrt{\frac{\eta\langle
P_{ab}\rangle}{4Bh\nu}} \delta 
\tilde{n}\sqrt{\frac{\sigma_{0}}{\tilde{n}}}
\end{equation}

Setting the SNR to unity gives the smallest measurable
change in column density from absorption:

\begin{equation}
    \label{eq:nminabthick}
\delta \tilde{n}(min)_{thick}=\sqrt{\frac{2.5Bh\nu}{\eta \langle
P_{ab}\rangle}\tilde{n}^{2}}
\end{equation}
\begin{equation}
    \label{eq:nminabthin}
\delta \tilde{n}(min)_{thin}=\sqrt{\frac{4Bh\nu}{\eta \langle
P_{ab}\rangle}\frac{\tilde{n}}{\sigma_{0}}}
\end{equation}

We might be more interested in fixing $\Gamma$, the photon absorption 
rate of an individual atom.  This can be found from the power 
absorbed by the condensate:
\begin{equation}
\frac{\langle P_{ab}\rangle}{h \nu}=\Gamma\;\tilde{n}\;A
\end{equation}
where $A$ is the cross-sectional area of the beam, and therefore 
$\tilde{n}A$ is the number
of atoms in the beam.  The smallest measurable change in column 
density is therefore given by

\begin{equation}
    \label{eq:nminabthickGamma}
\delta \tilde{n}(min)_{thick}=\sqrt{\frac{2.5\;B}{\eta \Gamma 
A}\tilde{n}}
\end{equation}
\begin{equation}
    \label{eq:nminabthinGamma}
\delta \tilde{n}(min)_{thin}=\sqrt{\frac{4\;B}{\eta \Gamma A 
\sigma_{0}}}
\end{equation}

\subsection{\label{sec:fl}Fluorescence}

The signal from fluorescence is the same as for absorption, with the
exception that only 1-10\% of the emitted photons would typically be
collected, reducing the signal by the collection efficiency 
$\Upsilon$.

The total photocurrent from the emitted photons collected on the
photodiode is:
\begin{equation}
    \label{eq:isigfl}
i=\rho \Upsilon P_{p0}(1-e^{-k_{p}}-e^{-k_{p}}\delta k_{p}
\sin(\omega_{p} t))
\end{equation}

As before, the desired signal is the AC component of $i$, which is 
selected
with a highpass filter on the current.  The current shot noise is 
related to the average of $i$, as described by (\ref{eq:shot}).  
The resulting RMS
signal and noise are given by:
\begin{eqnarray}
    \label{eq:sigfl}
i_{sig}&=&\frac{\rho \Upsilon}{\sqrt{2}}P_{p0}(e^{-k_{p}}\delta 
k_{p}) \\
    \label{eq:shotfl}
i_{shot}&=&\sqrt{2eB \rho \Upsilon P_{p0}(1-e^{-k_{p}})}
\end{eqnarray}

Taking the ratio of these results yields the SNR.  The non-destructive
criterion is
included by rewriting in terms of $\langle P_{ab} \rangle$.  Setting
$\delta k_{p}=\frac{\delta \tilde{n}}{\tilde{n}}\delta k_{p}$ gives:
\begin{equation}
\frac{S}{N}=\sqrt{\frac{ \Upsilon\eta\langle
P_{ab}\rangle}{4 B h\nu}}\frac{\delta \tilde{n}}{\tilde{n}}
\frac{e^{-k_{p}}k_{p}}{1-e^{-k_{p}}}
\end{equation}

This shows that the SNR for fluorescence will be at a maximum in the 
limit
of thin clouds, when $k_{p}\ll 1$.  For sensitivity to small changes 
in
the BEC, we would expect to detune to the linear thin cloud regime.
\begin{equation}
\frac{S}{N}=\sqrt{\frac{\Upsilon\eta\langle
P_{ab}\rangle}{4 Bh\nu}}\frac{\delta \tilde{n}}{\tilde{n}}
\end{equation}
The smallest measurable change in column density from fluorescence is
found by setting the SNR to unity.
\begin{equation}
    \label{eq:nminfl}
\delta \tilde{n}(min)=\sqrt{\frac{4Bh\nu}{\Upsilon \eta \langle
P_{ab}\rangle}\tilde{n}^{2}}
\end{equation}

In terms of the rate of absorption per atom, this is given by
\begin{equation}
    \label{eq:nminflGamma}
\delta \tilde{n}(min)=\sqrt{\frac{4B}{\Upsilon \eta \Gamma 
A}\tilde{n}}
\end{equation}

Fluorescence has the same maximum SNR (for a given $\langle
P_{ab}\rangle$) regardless of whether the BEC is optically thick or
thin to resonant light.  An important caveat is that in the optically
thick case, the reabsorption of emitted photons will usually require a
lower $\langle P_{ab}\rangle$ for the measurement to be
non-destructive.  

In the limit of an atomic cloud that is optically thick to 
resonant
light, the ratio of optimized SNL fluorescence to optimized SNL
absorption is approximately the collection efficiency.  In an actual
fluorescence experiment the collection efficiency will be much less
than one, and absorption is the better option for an optically thick
cloud.  In the optically thin limit, the ratio of optimized SNL
fluorescence to optimized SNL absorption equals 
$\sqrt{\frac{\sigma_{0}
\tilde{n}}{\Upsilon}}$.  Fluorescence is the most sensitive
technique in the case of very thin clouds, when 
$\sigma_{0}\tilde{n}\ll\Upsilon$.

\section{\label{sec:bigint}Dynamic non-destructive dispersive 
detection}

\subsection{\label{sec:int}Separated beam path interferometry}

\begin{figure}
\includegraphics{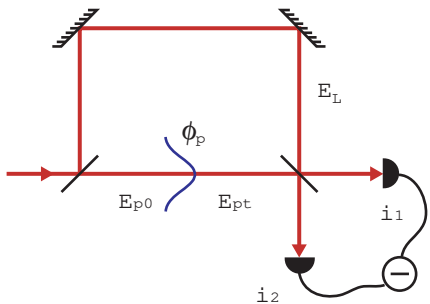}
\caption{\label{Fig:Figint1} A generic separated beam path
interferometer, where the local oscillator beam passes outside the
condensate.  A homodyne measurement is depicted, where the current
from the two ports of the interferometer are subtracted.}
\end{figure}

We analyze the generic separated beam path interferometer shown in 
Fig. 
\ref{Fig:Figint1}.  A local oscillator with power $P_{L} = \xi 
E_{L}^{2}$
passes outside the BEC, where $\xi$ is the proportionality constant
relating the square of the electric field to the power in the optical
field.  A probe with incident power $P_{p0}$ passes through the atoms,
experiencing the phase shift and absorption described in Sec. 
\ref{sec:ab}.

The current from the photodetectors, where + and $-$
refer to the two 
ports
of the interferometer, is:
\begin{equation}
\label{eq:isigint}
i_{1,2}=\frac{\rho}{2}(P_{L}+P_{pt}\pm 
2\sqrt{P_{L}P_{pt}}\cos(\phi_{tot}))
\end{equation}

The total phase shift, $\phi_{tot}$, is composed of the assumed stable
phase difference between the probe and local oscillator beams, 
$\phi_{lo}$,
and the AC and DC components of the phase shift from the BEC. The DC 
phase
shifts are combined by setting $\phi_{0}=\phi_{p}+\phi_{lo}$.  
\begin{eqnarray}
\label{eq:cosphi}
    \cos(\phi_{tot})&=&\cos(\phi_{0}+\delta
\phi_{p}\sin(\omega_{p}t))\nonumber\\
    &=&\cos\phi_{0}-\delta \phi_{p}\sin\phi_{0}\sin(\omega_{p}t)
\end{eqnarray}
In order to
maintain a constant operating point at $\phi_{0}$, the interferometer 
would
need to be locked at this point.  The bandwidth of the locking loop 
should
be sufficiently fast to track the slow decay, which for a typical BEC 
will
be at most $1$ Hz.  The bandwidth must also be much slower than the 
trap
oscillation frequency, where we expect our lowest signal frequencies.

We find the total photodetector currents by substituting equation
(\ref{eq:cosphi}) into (\ref{eq:isigint}).
\begin{eqnarray}
    \label{eq:isig1} i_{1,2}=\frac{\rho}{2}(P_{L}+P_{pt}\pm
    2\sqrt{P_{L}P_{p}} (\cos\phi_{0} \nonumber\\ -\delta
    \phi_{p}\sin\phi_{0}\sin(\omega_{p}t))
\end{eqnarray}

The best SNR is achieved with homodyne detection, detecting at both 
ports
and subtracting the currents, as will be shown in Section
\ref{sec:squeezing}.  The desired signal is the AC component of $i_{1}
- i_{2}$ which can be selected with an appropriate filter on the 
current. 
The resulting RMS signal, $\sqrt{\langle i_{ac}^{2}\rangle}$, is:
\begin{equation}
\label{eq:sigint}
i_{sig}=\rho \sqrt{2P_{L}P_{pt}}\delta \phi_{p}\sin\phi_{0}
\end{equation}

We assume the noise is dominated by shot noise from the laser, which 
is
related to current noise on the photodetector by Eq.  \ref{eq:shot}.  
The
current noise from each photodetector is added in quadrature:
\begin{eqnarray}
i_{shot}&=&\sqrt{2eB (\langle i_{1}\rangle+\langle i_{2}\rangle)}
\nonumber \\
\label{eq:shotint}
&=&\sqrt{2eB \rho (P_{L}+P_{pt})}
\end{eqnarray}

Taking the ratio of Eqs.  (\ref{eq:sigint}) and (\ref{eq:shotint}) 
gives
the SNR for a SNL interferometer:
\begin{equation}
    \label{eq:snint}
\frac{S}{N}=\sqrt{\frac{\eta}{Bh\nu} }\delta
\phi_{p} \sin\phi_{0}\sqrt{\frac{P_{L}P_{pt}}{P_{L}+P_{pt}}}
\end{equation}

As we will see later in the analysis, the power in the probe beam is
limited by the non-destructive requirement to a value far less than 
the
available laser power, even for large detunings.  By inspection, 
optimizing
equation (\ref{eq:snint}) subject to the restriction on the probe 
power
yields the optimum operating points of $P_{L}\gg P_{pt}$ and
$\phi_{0}=\frac{\pi}{2}$.  This operating point is halfway up a 
fringe,
with far greater power in the local oscillator than the probe beam.  
This
contrasts with an interferometer that is designed to measure a small 
phase
shift without this restriction, which is limited by the total 
available
laser power, and has optimal SNR when there is equal power in the two
interferometer paths.

At the optimum operating point, the SNR is
\begin{equation}
    \label{eq:largelo}
\frac{S}{N}=\sqrt{\frac{\eta}{Bh\nu}P_{pt}}\delta \phi_{p}
\end{equation}

Note that the optimum SNR is independent of power in the local 
oscillator
for a SNL measurement provided the shot noise dominates the detector 
noise 
in a real detector, and provided the power of the local oscillator is 
far 
greater than that of the probe.

We rewrite the SNR including the non-destructive criterion limiting 
absorption
from Eq. \ref{eq:nd}.  The details of the atom-light interaction are
included using Eqs.~(\ref{eq:ab}) and (\ref{eq:phase}).  We combine 
the
column density and absorption coefficient as the dimensionless 
variable
$\beta=\tilde{n}\sigma_{0}$.

\begin{equation}
    \label{eq:intfull}
\frac{S}{N}=\sqrt{\frac{\eta \langle P_{ab} \rangle}{4Bh\nu}}
\frac{\delta \tilde{n}}{\tilde{n}}
\frac{\frac{\beta \Delta }{1+\Delta^{2}}}
{\sqrt{\exp(\frac{\beta}{1+\Delta^{2}})-1}}
\end{equation}

\begin{figure}
\includegraphics{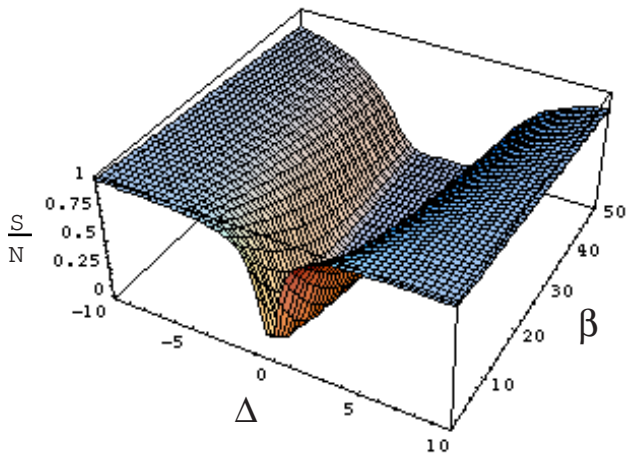}
\caption{\label{fig:figBa} The normalized signal to noise for
a SNL interferometer restricted by a non-destructive limit on the 
power
absorbed.  The SNR is a function of laser detuning, $\Delta$, in half
linewidths, and the optical thickness on resonance, $\beta $.  The SNR
reaches an optimum when the probe laser is sufficiently far detuned 
for the
BEC to become optically thin.  It is also necessary to be far 
detuned, even
when the BEC is optically thin on resonance, to be in the regime 
where the
dispersion scales as $\frac{1}{\Delta}$, and the absorption scales as
$\frac{1}{\Delta ^{2}}$.}
\end{figure}

The SNR is shown in Figure \ref{fig:figBa}, normalized to one.
The maximum occurs in the far detuned limit where the atomic
sample is optically thin.  These two limits are satisfied when 
$\Delta^{2}
\gg \beta + 1$.  In this limit, the optimum SNR is :
\begin{equation}
    \label{eq:thinsn}
\frac{S}{N}=\sqrt{\frac{\eta \langle P_{ab}\rangle}{4Bh\nu}
\frac{\sigma_{0}}{\tilde{n}}}\delta \tilde{n}
\end{equation}

Setting the SNR to unity gives the smallest measurable change in 
column
density from an optimized non-destructive interferometer.
\begin{equation}
    \label{eq:nminint}
\delta \tilde{n}(min)=\sqrt{\frac{4Bh\nu}{\eta \langle
P_{ab}\rangle}\frac{\tilde{n}}{\sigma_{0}}}
\end{equation}
In terms of the absorption rate per atom, this is given by
\begin{equation}
    \label{eq:nminintGamma}
\delta \tilde{n}(min)=\sqrt{\frac{4B}{\eta \Gamma A \sigma_0}}
\end{equation}

The SNR in an optimized interferometer is independent of the laser 
power
and detuning.  The smallest signal that can be detected depends only 
on the
column density of the BEC, the bandwidth of the measurement, and on 
the
stringency of the non-destructive criterion required for the 
particular
measurement.

In the limit of optically thin clouds on resonance
($\tilde{n}\sigma_{0}\ll1$), optimized interferometry has exactly the 
same
signal to noise as optimized absorption.  As the column density of 
the BEC
increases to the optically thick limit on resonance, the optimum SNR 
from
absorption drops by a factor of $\sqrt{0.6\tilde{n}\sigma_{0}}$, while
interferometry maintains the same maximum for both limits.  In this 
limit,
absorption has the same sensitivity as fluorescence (except for a 
factor of
collection efficiency).  Interferometry is fundamentally superior to
absorption or fluorescence for measuring BECs with 
$\tilde{n}\sigma_{0} \gg
1$\cite{Ketterle1999}.

\subsection{Frequency modulation spectroscopy.}

\begin{figure}
\includegraphics{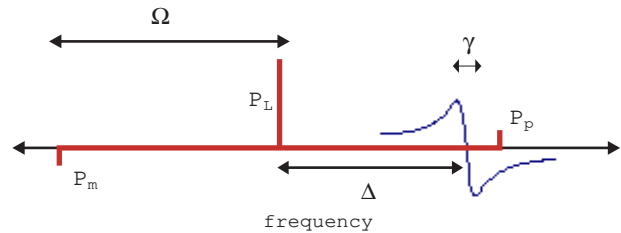}
\caption{\label{fig:figFMS} In frequency modulation spectroscopy, the
probe beam consists of a carrier with power, $P_{L}$, and two out of 
phase
sidebands at plus and minus the modulation frequency, $\Omega$, with 
power
$P_{m}$ and $P_{p}$ respectively.  The carrier is detuned $\Delta$ 
from the
atomic transition.  $\Delta$ and $\Omega$ are in units of half atomic
linewidths, $0.5 \gamma$.  The optimum detuning is found to be 
0.9$\Omega$,
with a sideband on each side of the atomic resonance as depicted in 
the
figure.}
\end{figure}

Frequency modulation spectroscopy (FMS), a single beam and therefore
geometrically stable technique, has previously been proposed by our 
group and 
the CNRS group as a non-destructive dynamic detector of BECs
\cite{Lye1999,Aspect1999}.  Instead of a separate local oscillator
that passes around the BEC, FMS relies on a frequency shifted local
oscillator, far detuned from resonance relative to the probe beam,
that passes through the BEC. In this section, we investigate the
important parameters in a FMS measurement, and consider whether it is
possible with available detectors to use FMS as a non-destructive 
probe for
a BEC feedback experiment.

In FMS, the single beam that passes through the BEC has a carrier, 
$P_{L}$,
at frequency, $\omega$, and two sidebands, $P_{p}$ and $P_{m}$, at
frequencies $\omega+\Omega$ and $\omega-\Omega$ respectively
\cite{Bjorklund1983} as shown in Fig.~\ref{fig:figFMS}.  The
signal, the sum of the two beat signals between the carrier and each 
of the
sidebands, is detected at the modulation frequency, $\Omega$.  With 
no BEC
present, there is zero signal as the two sidebands are out of phase 
and the
beats cancel.  With a BEC present, the three components of the beam 
receive
different phase shifts due to the frequency dependent dispersion of 
the
atoms described by $\phi_{L}$, $\phi_{p}$ and $\phi_{m}$, and the beat
signals no longer cancel.  In the limit of small phase shifts, the
amplitude of the net beat is proportional to the column density.  FMS 
not
only has the advantages of geometric stability, it has zero 
background thus
is insensitive to classical laser noise, and the large modulation 
frequency
enables detection in a quiet part of the laser intensity spectrum as 
well
as being far above any 1/f electronic noise.

After optimizing FMS in the following analysis, we find that even in 
the
best case scenario, the shot noise is well below the detector noise 
for a
typical PIN diode detector.  This best case occurs when we are 
detecting
the full BEC, $\delta \tilde{n} = \tilde{n}$ and we have the minimum
bandwidth possible to measure trap frequencies, B = 100 Hz.  In 
contrast to
all other techniques analyzed in this paper, we optimize the FMS signal
relative to detector noise in the following analysis.

The optical power incident on the photodetector is proportional to the 
square of the electric field averaged over an optical cycle:
\begin{eqnarray}
P_{opt} = \xi \langle \left(E_{L}\cos(\omega t+\phi_{L})+E_{p}\cos((\omega 
+\Omega
)t+\phi_{p})\nonumber\right. \\ \left.+E_{m}\cos((\omega-\Omega 
)t+\phi_{m})\right)^{2} \rangle
\end{eqnarray}

The incident optical power produces a current from the photodetector. 
Only the terms at the modulation frequency, $\Omega$, are of interest. 
We assume $P_{m}=P_{p}$, and write the sidebands as a (as yet
unspecified) fraction of the carrier power, $P_{p}=m^{2}P_{L}$.  We
rewrite the equation in terms of the total power,
$P_{tot}=P_{L}(1+2m^{2})$.

\begin{equation}
i_{sig}=2\rho \frac{mP_{tot}}{1+2m^{2}}(\cos(\Omega 
t+\phi_{L}-\phi_{m})-
\cos(\Omega t+\phi_{p}-\phi_{L}))
\end{equation}

The signal is mixed down to DC with a radio frequency local 
oscillator with
waveform $f(t)=\cos(\Omega t+\chi)$.  Assuming the mixer operates as
an ideal multiplier, the output current is $f(t)\times 
i_{sig}(t)$. 
As the gain is identical for signal and noise, we have set it 
equal to
one.  A lowpass filter is used to remove frequencies of $\Omega$ and 
above
to give an RMS voltage of:
\begin{eqnarray}
i_{sig}=\rho P_{tot}\frac{m}{1+2m^{2}}
\times ~~~~~~~~~~~~~~~~~~~~~~~~~~~~~~ \nonumber\\
\left(\cos(\phi_{L}-\phi_{m}-\chi) -
\cos(\phi_{p}-\phi_{L}-\chi)\right)
\end{eqnarray}

Maximum signal occurs when $\chi=\frac{\pi}{2}$.  We assume we are
operating at large laser detuning and that the phase shifts from the 
BEC
are small, $\phi_{m,p,L}\ll 1$.
\begin{equation}
i_{sig}=\rho P_{tot}\frac{m}{1+2m^{2}} (2\phi_{L}-\phi_{m} -\phi_{p})
\end{equation}

The phase shift dependence on detuning is included from equation
(\ref{eq:phase}), with detuning defined relative to the carrier 
frequency,
as shown in figure \ref{fig:figFMS}.  Again, we assume large 
detunings,
$\Delta\gg 1$.  The modulation frequency and the detuning is measured 
in
units of half atomic linewidths.
\begin{eqnarray}
i_{sig}=\rho P_{tot}\frac{m}{1+2m^{2}}
\times~~~~~~~~~~~~~~~ \nonumber\\
\frac{\tilde{n} \sigma_{0}}{2\Omega}\left(
\frac{2}{\Delta}-\frac{1}{\Delta+\Omega}
-\frac{1}{\Delta-\Omega}\right)
\end{eqnarray}

The signal is rewritten in terms of the ratio of the detuning to the
modulation frequency, $\frac{\Delta}{\Omega}=D$.
\begin{equation}
    \label{eq:sigFMS} i_{sig}=\rho P_{tot}\frac{m}{1+2m^{2}}
\frac{\tilde{n} \sigma_{0}}{\Omega}
\left(\frac{1}{D-D^{3}}\right)
\end{equation}

The sidebands and the carrier all contribute to absorption and all 
three
components must be included when fixing the average power absorbed to 
a
non-destructive level.
\begin{equation}
\langle P_{ab}\rangle=P_{tot}\langle k_{L}+m^{2}(k_{p}+k_{m}) \rangle
\end{equation}
The absorption dependence on detuning is included from equation
(\ref{eq:phase}).
\begin{equation}
\langle P_{ab}\rangle=P_{tot}\frac{\tilde{n} \sigma_{0}}{\Omega^{2}}
\left(\frac{1}{D^{2}}+\frac{m^{2}}{(D+1)^{2}}+\frac{m^{2}}{(D-1)^{2}}
\right)
\end{equation}

Substituting $\langle P_{ab} \rangle$ into equation (\ref{eq:sigFMS}):
\begin{equation}
   \label{eq:fullfms} i_{sig}=\frac{\rho \langle P_{ab} \rangle \Omega
\frac{m}{1+2m^{2}}}
{\left(D-D^{3}\right)
\left( \frac{1}{D^{2}}+\frac{m^{2}}{(D+1)^{2}}
+\frac{m^{2}}{(D-1)^{2}}\right)}
\end{equation}

\begin{figure}
\includegraphics{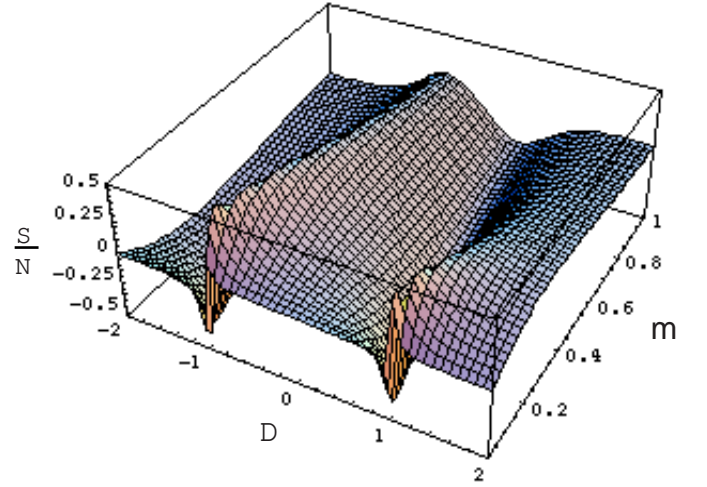}
\caption{\label{fig:figmvsR} The normalized signal to noise in a
detector limited non-destructive FMS measurement.  The SNR is a 
function of
the ratio of detuning to modulation frequency, $D$, and the ratio of 
power
in the carrier to power in the sidebands, $m^{2}$.  The optimum is at 
the
operating points $D=0.9$ and $m=0.1$.}
\end{figure}

The FMS signal from (\ref{eq:fullfms}), normalized by dividing 
through by
$\rho \langle P_{ab} \rangle \Omega $, is shown in figure
\ref{fig:figmvsR} versus $D$ and $m$.  The signal has an optimum
value of 0.5 at $D=0.9$ and $m=0.1$.  This is similar to the situation
shown in figure \ref{fig:figFMS}.
\begin{equation}
i_{sig}= 0.5 \rho \langle P_{ab} \rangle \Omega
\end{equation}

The optimum signal increases with modulation frequency which will be
limited by the bandwidth of the detector.  PIN diodes are the most 
suitable
detectors with their combination of high bandwidth and large dynamic 
range. 
The photodiode current will have noise at the modulation frequency,
$i_{n}$, which will be transferred through the mixer.
\begin{equation}
\frac{S}{N}=0.5\frac{\langle P_{ab} \rangle \Omega}
{NEP \sqrt{B}}
\end{equation}
Inserting the values $\Omega=\frac{15GHz}{0.5\gamma}=5000$, $\langle 
P_{ab}
\rangle=10^{-13}$W, $NEP=5\times 10^{-11}$, and B=100 Hz
\cite{NewFocus}:
\begin{equation}
\frac{S}{N}=1
\end{equation}

Unlike separated beam path interferometry, FMS improves in 
sensitivity with
increasing detuning and from atomic resonance and increasing 
modulation
frequency.  Although FMS offers the advantages of simplicity and
robustness, it is limited to relatively low 
local
oscillator powers by the speed of current detectors, and by 
the non-destructive criterion placed on the measurement.  It seems 
unlikely that it could be pushed into a 
SNL
regime.  Despite its suitability for many dynamic measurements, it 
would 
appear unlikely to be able to compete
with separated beam path interferometry in feedback applications.

\subsection{\label{sec:Cavity}Resonant Interferometry}

\begin{figure}
    \includegraphics{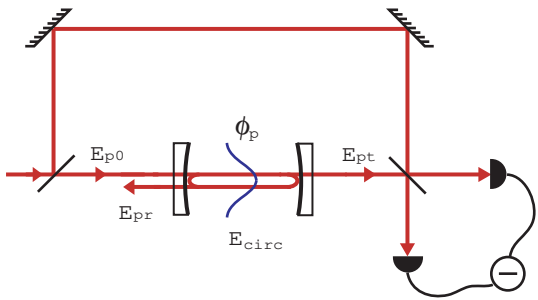}
\caption{\label{fig:cavity} A resonant interferometer.  A cavity is
placed in one arm of a generic interferometer, one of many equivalent
geometries utilizing a high finesse cavity to increase the 
sensitivity of a
non-destructive measurement.}
\end{figure}

In this section, we analyze a resonant cavity as a non-destructive 
detector
of Bose-Einstein condensates.  The cavity is included in one arm of a
Mach-Zender interferometer, as shown in Fig.~\ref{fig:cavity}, 
enabling an
interferometric homodyne measurement using the same optimum operating
points identified in the earlier analysis in Sec~\ref{sec:int}.  High 
finesse cavities
have been used in the strong coupling regime to both detect and control 
the motion of
single atoms \cite{RempeKimble}.  We are working in the weak coupling 
regime with the sole aim of extracting information about the condensate.

The ratio
of the amplitude of the reflected field to that of the field incident 
on a mirror is $r$.  Similarly, $t$ is the ratio of the amplitude of the
transmitted field to that of the incident field.  The single pass 
phase shift is denoted $\phi_{sp}$.  The reflected, transmitted and 
circulating electric fields of the cavity are given by the equations below.

\begin{eqnarray}
E_{r}&=&\frac{r 
E_{p0}\left(1-e^{2i\phi_{sp}}\right)}{1-r^{2}e^{2i\phi_{sp}}}
\nonumber \\ \nonumber \\
E_{t}&=&\frac{-t^{2}E_{p0}}{1-r^{2}e^{2i\phi_{sp}}}\nonumber 
\\ \nonumber \\
E_{circ}&=&\frac{itE_{p0}}{1-r^{2}e^{2i\phi_{sp}}}
\end{eqnarray}

In every pass through the cavity, some of the probe light will be 
absorbed 
by the BEC.  We assume that the finesse is dominated by losses at the 
mirrors rather than by losses in the BEC. That is, $k\ll\frac{1}{ 
{\cal F} }$ 
where ${\cal F} =\frac{1}{1-r^{2}}$ for a high finesse cavity.  
Although 
the power absorbed is fixed for a non-destructive measurement, the 
absorption coefficient is not, and can be reduced by increasing the 
detuning.  For a standard BEC, where $\tilde{n} \sigma_{0} = 300$, 
and 
at a maximum detuning of $10^{13}$ Hz, the finesse will be limited to 
$10^{9}$, much larger than any achievable experimental finesse.

The single pass phase shift consists of a DC phase shift from the 
BEC, a
small fluctuating phase shift from the BEC, and a phase shift from the
cavity itself.  The cavity is operated on resonance, with the 
combined DC
phase shifts locked to zero.  The phase shift of the probe beam is 
most
sensitive to changes in the BEC column density at this operating 
point.

The sum of the transmitted and reflected power at a mirror must equal 
the incident power, allowing us
to rewrite $-t^{2}=r^{2}-1$.  We assume that the fluctuating BEC 
phase is
very small.  The change in the transmitted field due to the 
fluctuations in the BEC is:

\begin{equation}
    E_{t}=E_{p0}e^{i 2 {\cal F} \delta \phi_{p} 
    sin(\omega_{p}t)}
\end{equation}
    
The phase shift from the BEC is increased by a factor of twice the 
finesse
compared to the phase shift in a non-resonant interferometer.  This 
extra
factor can be substituted directly into the optimized non-resonant
interferometer SNR, Eq.  (\ref{eq:largelo}).

\begin{eqnarray}
\frac{S}{N}=2\sqrt{\frac{\eta P_{p0}}{h\nu B}} {\cal F} \delta 
\phi_{p}  \\ \nonumber
\end{eqnarray}

The non-destructive limit on the power absorbed will depend on the
circulating power in the cavity, $\langle P_{ab} \rangle = P_{circ} 
k_{p}$
(for optically thin clouds).  On cavity resonance, the circulating
power is:
\begin{equation}
    P_{circ}= {\cal F} P_{p0}
\end{equation}

As we found earlier, the signal to noise is optimised when the probe 
beam
is far detuned from atomic resonance, and the BEC is optically thin.  
In
this limit, the atomic phase shift and absorption coefficient 
simplify to
$\delta \phi =\frac{\delta \tilde{n}\sigma_{0}}{2\Delta}$ and
$k=\frac{\tilde{n}\sigma_{0}}{\Delta^{2}}$.  Including the 
non-destructive
criterion and the details of the atom-light interactions in the SNR 
gives:

\begin{equation}
\frac{S}{N}=\sqrt{\frac{\eta P_{ab} {\cal F} \sigma_{0}}{h\nu B
\tilde{n}}} \delta \tilde{n}
\end{equation}
and
\begin{equation}
\delta \tilde{n}(min)=\sqrt{\frac{h\nu B
\tilde{n}}{\eta P_{ab} {\cal F} \sigma_{0}}}
\end{equation}
or, in terms of the atomic absorption rate,
\begin{equation}
\delta \tilde{n}(min)=\sqrt{\frac{B}{\eta \Gamma A {\cal F} 
\sigma_{0}}} 
\end{equation}

The sensitivity in a non-destructive resonant interferometric 
measurement is
enhanced by a factor of the square root of the finesse compared to a
non-resonant interferometer.

\section{\label{sec:squeezing}Quantum analysis of interferometric 
detection}

In this section we re-examine interferometric detection
using a quantized treatment of the light.  With the use of classical
light, 
non-destructive detection is ultimately limited by the shot noise of 
the detected
light.  This is an important practical limit, but it is not a
fundamental one that restricts all possible imaginable detection
schemes.  Although this limit can be pushed out by increasing 
the finesse of a resonant interferometer, there will be restrictions 
on the maximum usable finesse as we discuss in 
section~\ref{sec:technical}.  
We demonstrate that the limits 
to detection can be improved with the use of a non-classical light 
source, and show that in all cases
balanced homodyne detection, with a very small proportion of the laser
power going through the atomic cloud, will provide the best
SNR for a given absorption rate by the BEC.

\begin{figure}
\includegraphics{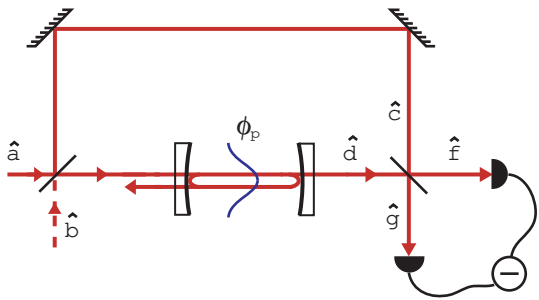}
\caption{\label{fig:sqzint}
A resonant interferometer with squeezed light,
$\hat{b}$, as one of the inputs.}
\end{figure}

Consider the interferometer described in section~\ref{sec:Cavity} with
squeezed light as one of the inputs.  We model this
interferometer with input fields described by the annihilation
operators $\hat{a}$ and $\hat{b}$ incident on a beam splitter with
reflectivity $R$. This setup is pictured in Fig. \ref{fig:sqzint}.

We assume that the field $\hat{a}$ is a large-amplitude coherent state
and that the field $\hat{b}$ is a (possibly squeezed) state with very 
low mean
amplitude.  The two input fields combine to give fields $\hat{c}$ and
$\hat{d}$ in the arms.  The field $\hat{d}$ experiences a phase
shift due to the atoms in the resonant cavity:
\begin{eqnarray}
\hat{c} &=& \sqrt{R} \hat{a}+i \sqrt{1-R}\hat{b}  \\
\hat{d} &=& (\sqrt{1-R} \hat{a}-i \sqrt{R}\hat{b})e^{i \phi} \nonumber
\end{eqnarray}
The phase difference $\phi$, in this discussion, contains both the 
systematic phase
shift between the probe and the local oscillator and the phase shift 
due 
to the atoms.  There are many equivalent energy-conserving choices for
these beamsplitter relations, but their only effect is to include an
extra constant phase in $\phi$.  The second beamsplitter recombines
these fields into two new fields $\hat{f}$ and $\hat{g}$:
\begin{eqnarray}
\hat{f} &=& \frac{1}{\sqrt{2}}( \hat{c}+i \hat{d})  \\
\hat{g} &=& \frac{1}{\sqrt{2}}( \hat{c}-i \hat{d}). \nonumber
\end{eqnarray}

The input field operators are approximated as the sum of a classical
coherent amplitude and a zero-mean quantum operator:
\begin{eqnarray}
\hat{a} = \alpha + \hat{\delta a}  \\
\hat{b} = \beta + \hat{\delta b} \nonumber
\end{eqnarray}
where $\hat{\delta a}$ and $\hat{\delta b}$  have the same commutation
relations as the input fields, $[\hat{\delta a},\hat{\delta
a}^{\dag}]=[\hat{\delta b},\hat{\delta b}^{\dag}]=1$.  We also
introduce the quadrature basis for the two input fields:
\begin{eqnarray}
\hat{\delta X_{j}^{+}} &=& \hat{\delta j}+\hat{\delta j}^{\dag}
 \\
\hat{\delta X_{j}^{-}} &=& i(\hat{\delta j}-\hat{\delta j}^{\dag}).
\nonumber
\end{eqnarray}
Keeping only terms
proportional to $\alpha$ or larger and assuming that the input field 
is a
strong local oscillator, we determine the number of photons at each 
detector.
$\hat{f}^{\dag}\hat{f}$ and $\hat{g}^{\dag}\hat{g}$, 
\begin{eqnarray}
\hat{f}^{\dag}\hat{f} &=& \frac{1}{2} \alpha 
\left\{(\alpha+\hat{\delta
X}_{a}^{+})
  \left(1-2\sqrt{(1-R)R}\sin(\phi)\right)\right. \nonumber \\
&&\left.+(2\beta+\hat{\delta
X}_{b}^{+})\cos(\phi)-(1-2R)\sin(\phi)\hat{\delta X}_b^{-}\right\}
 \\
\hat{g}^{\dag}\hat{g} &=& \frac{1}{2} \alpha 
\left\{(\alpha+\hat{\delta
X}_{a}^{+})
  \left(1+2\sqrt{(1-R)R}\sin(\phi)\right)\right.\nonumber\\
&&\left.-(2\beta+\hat{\delta
X}_{b}^{+})\cos(\phi)+(1-2R)\sin(\phi)\hat{\delta X}_b^{-}\right\}
\end{eqnarray}
These photon numbers are directly proportional to the photocurrents
received from the detectors, with the $\hat{\delta X}_j^{+,-}$
operators averaging to zero expectation value, but having non-zero
variance.  The product of the variances obey the inequality $\Delta
\hat{\delta X}_j^{+}\Delta \hat{\delta X}_j^{-}\ge 1$ due to
Heisenberg's Uncertainty Principle.  For shot-noise limited coherent
lasers, these terms each have a variance of unity.

Summing these intensities, we find that
\begin{equation}
\hat{f}^{\dag}\hat{f}+\hat{g}^{\dag}\hat{g}=\alpha^2+\alpha 
\hat{\delta
X}_a^+
\end{equation}
which is simply the intensity and shot noise of the input local
oscillator.

In order to find the ideal operating point for detection, we set
$\phi=\phi_0+\phi_p$.  $\phi_0$ is the systematic phase shift between
the fields $\hat{c}$ and $\hat{d}$, which depends on the phase
difference between the different paths of the interferometer and the 
cavity detuning, as well as the choice of beam-splitter relations.  
$\phi_{p}$ is the phase shift
induced by the atoms.  In the limit of small signal, $\phi_{p}\ll 1$, 
we can make the approximations $\sin(\phi) \approx
\sin(\phi_0)+\cos(\phi_0) \phi_p$ and $\cos(\phi) \approx
\cos(\phi_0)-\sin(\phi_0) \phi_p$, and readily extract the signal to
noise ratio.
\begin{eqnarray}
\hat{f}^{\dag}\hat{f} &=& \frac{1}{2} \alpha 
\left\{(2\beta+\hat{\delta
X}_{b}^{+})(\cos(\phi_0)-\sin(\phi_0) \phi_p) \right. \label{eq:fdagf}
\\
&&\left.-(1-2R)(\sin(\phi_0)+\cos(\phi_0) \phi_p)\hat{\delta
X}_b^{-}\right\}  \nonumber \\
&&\left.+(\alpha+\hat{\delta X}_{a}^{+}) ~~\times \right. \nonumber\\
&&\left(1-2\sqrt{(1-R)R}(\sin(\phi_0)+\cos(\phi_0) \phi_p)\right)
\nonumber \\
\hat{g}^{\dag}\hat{g} &=& \frac{1}{2} \alpha \left\{
(1-2R)(\sin(\phi_0)+\cos(\phi_0) \phi_p)\hat{\delta X}_b^{-}\right.
\label{eq:gdagg} \\
&&\left.
-(2\beta+\hat{\delta X}_{b}^{+})(\cos(\phi_0)-\sin(\phi_0) \phi_p)
\right. \nonumber \\
&&\left. +(\alpha+\hat{\delta X}_{a}^{+}) ~~\times \right. \nonumber\\
&&\left.\left(1+2\sqrt{(1-R)R}(\sin(\phi_0)+\cos(\phi_0)
\phi_p)\right)\right\}.  \nonumber
\end{eqnarray}

\subsection{Single port detection}

Assuming that the dominant noise source is the shot noise of the 
light,
and ignoring the finite quantum efficiency of the photodetectors, we 
can 
assume
that the photocurrent is directly proportional to the instantaneous
photon number with identical statistics.  We examine the signal
received from a single photodetector by taking the expectation value 
of
the number operator (either Eq.(\ref{eq:fdagf}) or
Eq.(\ref{eq:gdagg})).  We see that it has a constant component and a
signal proportional to $\phi_p$.  The noise is determined by examining
the variance of the number operator.  The three noise terms $\Delta
(\hat{X}_j^{+(-)})^2 = \langle (\hat{\delta X}_j^{+(-)})^2\rangle$ must be
added in quadrature, with $(\Delta \hat{X}_a^{+})^2=1$Hz for a shot-noise
limited input laser.  We ignore the DC component, which in practice
can be achieved either through modulation or by a difference
measurement.  For unit bandwidth, the signal to noise is given by
%
\begin{equation}
\left(\frac{S}{N}\right) = \frac{2\sqrt{(1-R)R} \;\alpha \cos(\phi_0)
\delta \phi_{p}}
{\sqrt{V}} \label{eq:stononeport}
\end{equation}
where
\begin{eqnarray}
V&=&\left(1-2\sqrt{(1-R)R}(\delta{\phi_{p}} 
\cos(\phi_0)+\sin(\phi_0))\right)^2 {\Delta \hat{X}_a^{+}}^2 \nonumber \\
&&+(1-2R)^2(\sin(\phi_0)+\cos(\phi_0) \delta\phi_{p})^2 {\Delta 
\hat{X}_b^{-}}^2 \nonumber \\
&&+{\Delta \hat{X}_{b}^{+}}^2(\cos(\phi_0)-\sin(\phi_0) \delta\phi_{p})^2
\end{eqnarray}
%
 
This has a maximum value for $R=1/2$ operating near a dark port.  The 
restriction on absorption by the BEC has not been included. We are guided 
by 
the fact that 
$\sqrt{1-R}\; \alpha$ is equal to the square root of the photon flux 
incident on the cavity.  Following section~(\ref{sec:Cavity}) and 
using equations~(\ref{eq:ab}),(\ref{eq:phase}) and (\ref{eq:nd}), we 
can make the substitution:
\begin{eqnarray}
\sqrt{1-R}\; \alpha \;\delta\phi_{p} &=&\sqrt{\frac{P_{p0}}{h \nu}} 
\;\delta\phi_{p} \nonumber\\
&=&\sqrt{\frac{P_{circ}}{{\cal F} h \nu}} \frac{{\cal F} \;\delta 
\tilde{n} \;\sigma_0}{2 \Delta} \nonumber\\
&=& \frac{\delta\tilde{n}}{2} \; \sqrt{A\;\sigma_0 \;{\cal 
F}\;\Gamma}   \label{eq:gammasubs}
\end{eqnarray}
where $\Gamma$ is the spontaneous emission rate of the atoms caused 
by the 
absorption, $A$ is
the cross-sectional area of the beam, and $\sigma_0=3\lambda^2/(2 
\pi)$ is the atomic cross-section as previously defined.  Using this
substitution in Eq.(\ref{eq:stononeport}) and maximizing the SNR for
choice of $\phi_0$ and $R$, we find that in the presence of squeezed 
light, 
the best operating parameters appear to be unchanged, but this 
corresponds to 
such a weak beam entering the interferometer that the signal will be 
dominated by detector noise rather than the shot
noise of the light.  This means that squeezed light will not usefully 
enhance the SNR for single port detection.  This is not the case 
for systems without a non-destructive criterion, as they do not have 
a cap
on the total circulating power in the cavity. 

In the absence of squeezing, the SNR for unit bandwidth is given by
\begin{equation}
\frac{S}{N} = \tilde{\delta n} \sqrt{\frac{A \;\sigma_0\; {\cal 
F}\;\Gamma}
{2}}.   \label{eq:stononeportopt}
\end{equation}

This limiting SNR assumes the detector has perfect efficiency and the 
absorption by the condensate is negligible. This is valid in the high 
finesse and high detuning limits.

\subsection{Homodyne measurement}

When an interferometer is used to measure a phase shift, it is
clearly wise to measure a large phase shift if this can be arranged.  
The 
phase shift 
from the BEC is
proportional to the square root of the absorption
rate, which is the main measure of the "destructiveness", and our 
measurement 
is constrained
in that we must detect as small a phase shift as possible. 
Without this constraint, 
single port detection has the
same theoretical maximum SNR as a homodyne measurement.  Including the
constraint, we find that this is no longer true.  A better maximum SNR
can be obtained by detecting both output ports of the interferometer
and examining the difference photocurrent, removing
the component of the shot noise which is correlated on each port.  In
operator form, the photon difference is:
\begin{eqnarray}
\hat{g}^{\dag}\hat{g} - \hat{f}^{\dag}\hat{f} =
(1-2R)\;\alpha \;\hat{\delta X}_b^{-} \;(\sin(\phi_0)+\cos(\phi_0)
\phi_p)  \label{eq:diff} \\
-(2\beta+\hat{\delta X}_{b}^{+})\alpha(\cos(\phi_0)-\sin(\phi_0)
\phi_p)  \nonumber \\
+2(\alpha^2+\hat{\delta X}_{a}^{+}\alpha)
\sqrt{(1-R)R}(\sin(\phi_0)+\cos(\phi_0) \phi_p).  \nonumber
\end{eqnarray}
The signal and noises are determined in the same manner as single
port detection discussed above.  After substituting the expression in
Eq.({\ref{eq:gammasubs}) for $\overline{\phi_p}$ to determine the SNR 
as a
function of atomic spontaneous emission rate $\Gamma$, we find that for a 
unit bandwidth it is given by
\begin{widetext}
\begin{equation}
\frac{S}{N} = \frac{\sqrt{R \;A \;\sigma_0 \;{\cal F} \;\Gamma} \; 
\cos(\phi_0) \;\delta 
\tilde{n}}
{\sqrt{4(1-R)R(\phi_p \cos(\phi_0)+\sin(\phi_0))^2+
(1-2R)^2(\sin(\phi_0)+\cos(\phi_0) \phi_p)^2 {\Delta \hat{X}_b^{-}}^2+
{\Delta \hat{X}_{b}^{+}}^2(\cos(\phi_0)-\sin(\phi_0) \phi_p)^2}}.
\label{eq:stondiff}
\end{equation}
\end{widetext}

This SNR is maximal when $\phi_0=0$ and $(1-R)\ll 1$, 
corresponding
to balanced detection with only a very small fraction of the input
laser power going through the atomic sample.  In this limit, we find
that the SNR has a theoretical maximum:
\begin{equation}
\frac{S}{N} = \frac{\delta \tilde{n} \;\sqrt{A \;\sigma_0\;{\cal 
F}\;\Gamma}}
{\Delta \hat{X}_{b}^{+}}.   \label{eq:stondiffopt}
\end{equation}
which, by comparison with Eq.(\ref{eq:stononeportopt}), is 
a factor of $\sqrt{2}$ larger than the optimal result for a single 
port
detection.  Both of these results show that with squeezing of the
appropriate quadrature of the vacuum input to the interferometer, an
arbitrarily high SNR can be achieved independent of the finesse of
the cavity.  Such an experiment would be difficult to demonstrate. It 
is only 
recently that squeezing has been used to improve the sensitivity of 
any
interferometer \cite{Mckenzie2002}.

\section{\label{sec:technical}Comparison of techniques and technical 
limitations in real 
detectors}

\subsection{Comparison of shot-noise limited techniques}

\begin{table}
\caption{\label{tab:comp}Smallest measurable change in column
density for the different techniques in a shot-noise limited 
measurement. 
The use of non-classical light would improve the sensitivity of all 
the
techniques by the squeezing factor.  Calculation of $\langle 
P_{ab}\rangle$
must take into account reabsorption in the optically thick limit.}
\begin{ruledtabular}
\begin{tabular}{l | c | c}
&&\\
Measurement Scheme & $\delta \tilde{n}(min)$ \;\;($P_{ab}$) & $\delta 
\tilde{n}(min)$\;\;($\Gamma$)\\
&&\\
\hline &&\\
Fluorescence&$\sqrt{\frac{4Bh\nu}{\Upsilon\eta \langle
P_{ab}\rangle}\tilde{n}^{2}}$&
$\sqrt{\frac{4B}{\Upsilon\eta \Gamma A}\tilde{n}}$\\ &&\\
Absorption (thick)& $\sqrt{\frac{2.5Bh\nu}{\eta \langle
P_{ab}\rangle}\tilde{n}^{2}} $&
$\sqrt{\frac{2.5B}{\eta \Gamma A}\tilde{n}} $\\ &&\\
Absorption (thin)& $\sqrt{\frac{4Bh\nu}{\eta \langle
P_{ab}\rangle}\frac{\tilde{n}}{\sigma_{0}}}$&
$\sqrt{\frac{4B}{\eta \Gamma A \sigma_{0}}}$\\ &&\\
Interferometry&$\sqrt{\frac{4Bh\nu}{\eta \langle
P_{ab}\rangle}\frac{\tilde{n}}{\sigma_{0}}}$&
$\sqrt{\frac{4B}{\eta \Gamma A \sigma_{0}}}$\\ &&\\
Resonant Interferometry&$\sqrt{\frac{Bh\nu}{ {\cal F} \eta  \langle
P_{ab}\rangle}\frac{\tilde{n}}{\sigma_{0}}}$&
$\sqrt{\frac{B}{ {\cal F} \eta \Gamma A \sigma_{0}}}$\\ &&\\
\end{tabular}
\end{ruledtabular}
\end{table}

Table \ref{tab:comp} shows a summary of the smallest measurable change
in column density using optimized absorption, fluorescence,
interferometry, and resonant interferometry in a shot-noise limited
measurement.  Fluorescence, interferometry, and resonant
interferometry have the same optimum operating point for all BEC
column densities, while absorption has the added complication that its
optimum changes depending on whether the BEC is optically thick or
thin when probed with resonant light.  In the limit of an optically
thick cloud, optimized absorption has the same sensitivity as
fluorescence, except for the factor of collection efficiency.  However
in this limit, interferometry or resonant interferometry are the
superior detectors to either absorption or fluorescence.  In the limit
of an optically thin cloud, the optimum absorption sensitivity is the
same as for non-resonant interferometry, but in this limit
fluorescence is now the most sensitive technique.  Either way,
absorption is never the most sensitive technique, and at best is
equally sensitive for a small range of measurements where $\tilde{n}
\sigma_{0}$ is slightly less than one.  Resonant and non-resonant
interferometry scale the same with respect to column density, however
a resonant interferometer is a factor of $\sqrt{ {\cal F} }$ more
sensitive.  The ratio of minimum $\delta \tilde{n}$ from resonant
interferometry and fluorescence shows how the ideal technique depends
on the BEC column density:
\begin{equation}
\frac{\delta \tilde{n}_{fl}}{\delta
\tilde{n}_{resint}}=\sqrt{\frac{4{\cal F} 
\sigma_{0}\tilde{n}}{\Upsilon}}
\end{equation}

In the limit of very thin clouds, $4 {\cal F} \sigma_{0}\tilde{n} \ll
\Upsilon$, fluorescence is the most sensitive technique, otherwise
resonant interferometry has the highest fundamental sensitivity.  
Resonant
interferometry is the only technique that can, at least 
theoretically, be
improved arbitrarily for fixed absorbed power.  The sensitivity of the
other techniques are limited by the experimental requirements on
non-destructiveness, bandwidth, and the column density of the BEC.

\subsection{Technical limitations with real detectors}

We have assumed in all but one of the detection schemes presented,
that laser shot noise dominates all other noise sources.  Shot-noise
limited sources and detection at the shot noise limit is standard
technology in quantum optics labs around the world.  Of more serious
concern in interferometric measurements are geometrical phase shifts
brought about by the acoustic vibration of beam splitters and mirrors. 
Here FMS, as a single beam method, is superior to all other
interferometric techniques.  In FMS, however, the local oscillator
passes through the BEC and it is far more destructive than separated
beam path interferometry.  It is in the design of separated beam path
methods and resonant interferometry that geometric phase shifts must
be carefully considered.

In Section \ref{sec:int}, it was shown that the signal to noise in a
separated beam path interferometer is independent of laser detuning
provided the detuning is sufficiently large that the cloud is
optically thin.  In addition to this restriction, we must avoid
lensing of the light beam due to the condensate.  Both can be achieved
by operating at small phase shifts.  The present model, although it
provides scaling and best case signal to noise, completely neglects
lensing effects on the propagation of the Gaussian probe beam. 
Excessive lensing will make signals difficult if not impossible to
interpret due to imperfect mode matching in separated beam path
interferometers and multimode behaviour in a resonant cavity.  The
effect will be particularly acute in a high finesse cavity and will
probably limit the maximum useful finesse.  Nonetheless,
interferometric measurements will be least sensitive to vibration if
the phase shift from the condensate is made as large as possible by
operating as close to atomic resonance consistent with the
restrictions above.  This highlights an important difference between
many interferometric measurements where small phase shifts are
detected with high intra-cavity power, and a non-destructive BEC
measurement where low probe powers are used and comparatively large
phase shifts can be detected.

In addition to the choice of detuning, a sensitive interferometer will
need to be acoustically and vibrationally isolated and have its
operating point locked in order to minimize geometrical phase shifts. 
There is a wealth of information in the literature on locking, and we
discuss here only a few points pertinent to measurements on BECs. 
Unlike many phase objects, atoms are a resonant system, and this can
be used to advantage.  Two phase coherent probe beams can be injected
into the interferometer with different detunings from atomic
resonance.  The beam closer to resonance will carry more information
on the condensate and less on geometric shifts.  The reverse is true
for the beam detuned further from resonance.  Comparison of the two
signals will allow locking of the operating point across the entire
signal band providing greater immunity to vibration than would be
possible on measurements of a non-resonant phase object.

Although the high finesse cavity is more sensitive than non-resonant
interferometry by a factor of the square root of the finesse, this
sensitivity comes at the price of increased susceptibility to
vibration.  Geometrical instabilities are amplified by the finesse.  A
high degree of vibration isolation, locking over the entire signal
bandwidth, and monolithic construction are probably essential if the
advantages of the high finesse cavity are to be realized.  This is not
an easy detector to build.  For many measurements, the relative
simplicity of the non-resonant interferometer may swing the balance in
its favour.  

A common approach to measuring a small phase shift in a
cavity is the Pound Drever Hall (PDH) method, whereby frequency modulated
light is injected into the cavity and the beat between the reflected
sidebands and the carrier are measured on a fast photodiode.  With the
carrier resonant, the beat signal is proportional to the phase shift. 
A quick calculation suggests this method is unsuitable for measuring
the phase shift induced by a BEC. Even at maximum detuning from atomic
resonance, the largest circulating power we could tolerate is 1 mW.
With a finesse of $10^{4}$, the input carrier power would be 100 nW.
Assuming we use standard modulation techniques, the maximum power in
the sidebands would be on the order of 10nW. As the sidebands are the
local oscillator in this measurement, it would seem unrealistic to
make the measurement shot-noise limited with a high bandwidth
detector.  An alternative but related technique is to make an
off-resonant PDH measurement.  Here, the carrier is detuned from both
the atomic resonance and the cavity resonance.  The carrier is
reflected from the cavity and provides a strong local oscillator.  The
modulation frequency of the input beam is matched to the cavity free
spectral range, and it is now the sidebands that circulate in the
cavity and probe the BEC. A similar signal to noise is obtained by
detecting the transmitted beam but operating off cavity resonance such
that the transmitted power is half of the input power.  This is
equivalent to operating half way down a bright fringe in an
interferometer.  The advantage of this technique is experimental
simplicity.  The disadvantage is an increased susceptibility to classical 
laser noise.

For sensitive detection, there are three basic photo-detector choices:
PIN photodiodes, avalanche photodiodes(APD), and photomultiplier
tubes(PMT) \cite{Donati2000}.  Both APD (in Geiger mode) and PMT are
single photon counters.  They are however limited to low photon flux
and typically can detect a maximum flux of $10^{6}$-$10^{7}$
photons/sec.  This limits the maximum transmitted power to $10^{-13}
$W limiting the bandwidth and prohibiting modulation at frequencies
high enough to avoid typical laser relaxation oscillations.  Detectors
based on PIN diodes designed for SNL measurements, have a dynamic
range on the order of $10^9$ and bandwidths of 1-10 GHz. Such detectors
are capable of handling large photon fluxes, high modulation
frequencies and operating at the shot noise limit. 
\cite{Gray1997,NewFocus}.

An upper limit on the power absorbed by the BEC during a
non-destructive measurement can be estimated by requiring that the
atom loss rate in the absence of the probe beam is equal to the atom
loss rate due to the probe beam.  We assume that one photon absorbed
corresponds to one atom lost.  This assumption does not take into
account reabsorption of photons in optically thick clouds, or the
effects of heating if the atom does not immediately leave the trap
after absorption of a photon.  Both these effects, however, make the
non-destructive criterion more stringent.  For a BEC with $10^{6}$
atoms and a lifetime of 1 second, this leads to an upper limit
$\langle P_{ab}\rangle = 10^{-13}$ W. With this power, the signal from
absorption and fluorescence will never be above the NEP of a PIN
diode.  Non-destructive, dynamic measurements of absorption or
fluorescence are restricted to APDs or PMTs.  Interferometry has the
option of using APDs, PMTs or PIN diodes.  The latter have sufficient
bandwidth for the modulation that will be required if we are to apply
standard squeezing techniques to these measurements.

\section{Conclusion}

With a few exceptions, the vast amount of information on BECs that has
been gathered in the last eight years has been recorded using CCD
cameras.  Although quiet, these detectors are slow and not suited to
dynamic detection and feedback.  The dynamic detection of condensates
described in this paper will be required if we are to use feedback to
reduce quantum noise on an atom laser beam.  Although alternative
detection schemes appear feasible for metastable helium, there are
advantages to rubidium \cite{Ian}.

In this paper we have proposed and analyzed a series of non-destructive
measurement schemes for atomic clouds.  The most sensitive is a new
proposal based on an optical cavity within an interferometer, although
it would be the hardest to implement in practice.  We contrast the
performance of this detector with a variety of dynamic detection
schemes for Bose-Einstein condensates based on interferometry,
fluorescence and absorption.  When these schemes are optimized subject
to fixed heating, we find that resonant interferometry is the only
scheme which can achieve an arbitrarily high SNR.

We find that for separated beam path interferometers, where the local
oscillator passes around the BEC and does not contribute to heating,
the signal to noise cannot be increased arbitrarily by detuning from
atomic resonance and increasing laser power.  For interferometric
techniques such as frequency modulation spectroscopy, where the local
oscillator passes through the BEC, the signal to noise can be improved
by detuning and increasing power but it will only ever approach the
SNR of the shot-noise limited separated beam path interferometer.  The
limitation of FMS is that the SNR is maximized where the modulation
frequency is of the same order of the detuning.  Available detectors
limit this to roughly 10 GHz.

Although resonant interferometry can be arbitrarily increased through
increasing the finesse, the SNR will ultimately be limited by the tight
experimental requirements which encumber a high finesse cavity.  For
many measurements, it may be preferable to use one of the simpler, 
less sensitive,
techniques.  In the optically thick regime, interferometry has 
greater sensitivity than either fluorescence or absorption.
In the thin regime, fluorescence is more
sensitive.  Absorption is the least sensitive in all circumstances.

The schemes we have presented can be used to detect classical
oscillations if the probe beam is focussed to a waist smaller than the
condensate.  Such a design, with feedback to the trap, could be used
either to enforce single mode operation or to mode-lock an atom laser
in order to provide a pulsed output.  Alternatively, if the probe beam
is larger than the condensate, these detectors can provide information
on number fluctuations.  This can be used to minimize the linewidth of
a pumped atom laser.  Spatial information on a condensate could be
obtained by scanning the probe beam in one or two dimensions using
acousto-optic modulators or micro-electronic mirrors.  Provided the
scan rate is significantly higher than all signal fluctuations of
interest, dynamic spatial information can be extracted and fed back to
the condensate in real time.  Although this may be difficult (but
certainly not impossible) in a separated beam path interferometer, it
would be relatively straight-forward to implement with fluorescence,
absorption or frequency modulation spectroscopy.

The fast photodiodes that are the basis of the techniques we have
described here are consistent with the future implementation of 
squeezed
light to improve the signal to noise. Although the gains that could be
made with present levels of squeezing are not great, this may become 
relevant 
as squeezing improves. We are currently designing detection of atoms 
on a 
chip based
on the techniques we have described here. These designs are compatible
with microchip BECs. The future implementation of microchip BECs with 
on
board non-destructive detection using squeezed light is an exciting
possibility.  If implemented with a split photodiode, these schemes 
are
compatible with sub diffraction limited resolution through spatial
squeezing.

Although light has many advantages, inherent non-linearity and the 
finite
rest mass of atoms promise benefits in many applications
\cite{Kasevich,Gupta2002}.  For this reason, the development of the 
pumped 
atom laser is an outstanding goal in atom optics. Although there have 
been
some early experiments in this field, the development of the pumped 
atom
laser will truly usher in the age of quantum atom optics. Initial
investigations indicate that pumping either by forced evaporation or 
by
spontaneous emission from an excited state may only produce a stable 
BEC under
particular conditions of density, temperature and scattering
length \cite{Haine2002}.  Stability of atom laser sources may be 
expected to 
improve dramatically if feedback techniques can be employed.
While there has been significant progress
in atom lasers over the last few years, dynamic detectors for
quantum atom optics experiments have not yet been developed. In this
paper, we have outlined the design criteria for dynamic atom detectors
based on single photon scattering. The experimental realization of 
these
detectors, their performance and the implementation of feedback will 
the
subject of future papers from our group.

\acknowledgments

This work was conducted in an Australian Research Council Centre of 
Excellence.  We thank Nick Robins for providing Figure 1.  
We would also like to thank Nicolas Treps, Ping Koy Lam, Ben 
Buchler, Stan Whitcomb, Daniel Shaddock, Bram Slagmolen, Malcolm Gray 
and Craig Savage for 
their valuable discussions.


\begin{thebibliography}{9}


\bibitem{Ketterle1996}
 M. R. Andrews {\it et al.}, Science {\bf 273}, 84-87 (1996)
\bibitem{Ketterle1999}W.
 Ketterle, D. S. Durfee and D. M. Stamper-Kurn, in "Bose-Einstein
 Condensation in Atomic Gases" , Proceedings of the International
 School of Physics, IOS Press 67-164 (1999).
\bibitem{Lye1999} J. E. Lye {\it et al.}, J. Opt.  B: Quantum 
Semiclassical Opt.  {\bf
 1}, 402-407 (1999).
\bibitem{Aspect1999}
V. Savalli {\it et al.}, Opt.  Lett.  {\bf 24}, 1552
- 1554 (1999).
\bibitem{Hulet1997} C. C. Bradley , C. A. Sackett, and
R. G. Hulet , Phys.  Rev.  Lett.  {\bf 78}, 985 - 989 (1997).
\bibitem{Milburn1998} J. F. Corney and G. J. Milburn, Phys.  Rev.  A.
{\bf 58}, 2399 - 2406 (1998).
\bibitem{Walker2001}S. Kadlecek, J.
Sebby, R. Newell, and T. G. Walker, Opt.  Lett.  {\bf 26}, 137 - 139
(2001).
\bibitem{Thomsen2001} H. M. Wiseman and L. K. Thomsen, Phys.  Rev.
Lett.  {\bf 86}, 1143 (2001).
\bibitem{Wiseman2002} L. K. Thomsen and
H. M. Wiseman, Phys.  Rev.  A {\bf 65}, 063607-1 (2002).
\bibitem{Treps2002} N. Treps {\it et al.}, Phys.  Rev.  Lett.  {\bf
88}, 203601 (2002).
\bibitem{Mckenzie2002} Kirk McKenzie {\it et al.}, Phys.  Rev.  Lett.  {\bf
88}, 231102 (2002).
\bibitem{laser1}M. R. Andrews {\it et al.}, Science {\bf 275}, 637
(1997).
\bibitem{laser2} E. W. Hagley {\it et al.}, Science,
{\bf 283}, 1706 (1999).
\bibitem{laser3} I. Bloch, T. W. Hansch, T.
Esslinger, Phys.  Rev.  Lett, {\bf 82}, 3008 (1999).
\bibitem{laser4}
J. L. Martin et.  al., J. Phys.  B: At Mol.  Opt.  Phys.  {\bf 32},
3065 (1999).
\bibitem{laser5} B. P. Anderson and M. A. Kasevich,
Science, {\bf 282}, 1686 (1998).
\bibitem{Esslinger2002} M. Kohl {\it
et al.} Phys.  Rev.  A {\bf 65}, 021606 (2002).
\bibitem{Leduc2001}
F. Pereira Dos Santos {\it et al.}, Phys.  Rev.  Lett.  {\bf 86}, 3459 
(2001).
\bibitem{Aspect2001} A. Robert {\it et al.}, Science 292, 461
(2001).
\bibitem{Gray1995} Quantum Noise Limited Interferometry, M. B.
Gray, PhD thesis, Australian National University (1995).
\bibitem{Caves1981} C. M. Caves, Phys.  Rev.  D. {\bf 23}, 1693 (1981).
\bibitem{Loudon1981} R. Loudon Phys.  Rev.  Lett.  {\bf 47} 815 (1981).
\bibitem{McClelland1993} A. J. Stevenson {\it et al.}, Appl.  Opt.  {\bf 
32} 3481 (1993).
\bibitem{Yariv1985}
A. Yariv, Optical Electronics, chapter 10, 3rd edition, CBS College
Publishing, New York, (1985).
\bibitem{RempeKimble}T. Fischer {\it et al.}, \prl {\bf 88}, 163002 
(2002); J. Ye, D.W. Vernooy and H.J. Kimble, \prl {\bf 83}, 4987 (1999).
\bibitem{Donati2000} S. Donati,
Photodetectors, Devices, Circuits and Applications, Prentice Hall New
Jersey (2000).
\bibitem{Robins2001}
N. Robins, C. Savage, and E. A. Ostrovskaya
Phys. Rev. A 64, 043605 (2001)
\bibitem{Gray1997} M. B. Gray {\it et al.} Rev. Sci. Instrum. {\bf 
69} 3755 (1998).
\bibitem{NewFocus}  Fast PIN photodiode, part no : 1480-S , {\it
www.newfocus.com.au}
\bibitem{Bjorklund1983} G. C. Bjorklund and M. D. Levenson, Appl.
Phys. B {\bf 32}, 145-152 (1983).
\bibitem{Ian}There appear to be no Feshbach resonances for metastable 
helium. (Ian Whittingham, private communication).
\bibitem{Haine2002} S. A. Haine, J. J. Hope, N. P. Robins, and C. M.
Savage, Phys. Rev. Lett. {\bf 88} 170403-1 (2002).
\bibitem{Kasevich}J.M. McGuirk, M.J. Snadden and M.A. Kasevich, \prl {\bf 
85}, 4498 (2000).
\bibitem{Gupta2002}S. Gupta {\it et al.}, \prl {\bf 89}, 140401 (2002).

\end{thebibliography}
\end{document}